\documentclass[3p,times,twocolumn]{elsarticle}

\usepackage{ecrcfake}
\usepackage{upgreek}

\volume{00}

\firstpage{1}

\journalname{}

\runauth{L. Tibaldo, I. A. Grenier et al.}

\jid{nuphbp}

\jnltitlelogo{Nuclear Physics B Proceedings Supplement}

\usepackage{amssymb}
\newcommand{\F}{\textit{Fermi}}
\newcommand{\g}{$\gamma$}

\newcommand{\hi}{\mathrm{H\,\scriptstyle{I}}}

\newcommand{\hd}{\mathrm{H}_2}

\newcommand{\apj}{Astrophys. J. }
\newcommand{\apjl}{Astrophys. J. Lett. }
\newcommand{\apjs}{Astrophys. J. Suppl. S. }

\newcommand{\nat}{Nature }
\newcommand{\aap}{Astron. Astrophys. }

\newcommand{\ssr}{Space Sci. Rev. }

\begin{document}

\begin{frontmatter}



%% \dochead{}

\title{The \F~LAT view of Cygnus: a laboratory to understand cosmic-ray acceleration and transport}


\author[slac]{L.~Tibaldo\fnref{fn1}}
\ead{ltibaldo@slac.stanford.edu}

\author[saclay]{I.A.~Grenier\fnref{fn2}}
\ead{isabelle.grenier@cea.fr}

\author{on behalf of the \F~LAT collaboration}

\address[slac]{KIPAC -- SLAC National Accelerator Laboratory, 2575 Sand Hill Rd. MS~29, Menlo Park,
CA 94025, USA}

\address[saclay]{Laboratoire AIM, CEA-IRFU/CNRS/Universit\'e Paris Diderot, Service d'Astrophysique,
CEA Saclay, 91191 Gif sur Yvette, France.}

\fntext[fn1]{Formerly: Dipartimento di Fisica e Astronomia ``G. Galilei'', Universit\`a di Padova
and INFN -- Sezione di Padova, Italy.}
\fntext[fn2]{Member of Institut Universitaire de France.}

\begin{abstract}
Cygnus~X is a conspicuous massive-star forming region in the Local Arm of the
Galaxy at $\sim
1.4$~kpc from the solar system. \g-ray observations can be used to trace cosmic rays (CRs)
interacting with the ambient interstellar gas and low-energy radiation
fields. Using the Fermi Large Area
Telescope (LAT) we have discovered the presence of a 50-pc wide cocoon of freshly-accelerated CRs in
the region bounded by the ionization fronts from the young stellar clusters. On the other hand, the
LAT data show that the CR population averaged over the whole Cygnus complex on a scale of $\sim
400$~pc is similar to that found in the interstellar space near the Sun. These
results confirm the long-standing hypothesis that massive star-forming regions host CR
factories and shed a new light on the early phases of CR life in such a turbulent
environment.
\end{abstract}


\end{frontmatter}


\section{Introduction}

Cosmic rays (CRs) are one of the most intriguing puzzles of the last century Physics.
It is strongly advocated that Galactic CRs are accelerated by supernova remnants (SNRs), which are
energetic and numerous enough to sustain
the CR population directly measured near the Earth \cite{ginzburg1964}. A
plausible mechanism
for CR acceleration in SNRs is provided by non-linear diffusive shock
acceleration.
The presence of high-energy electrons in SNRs is undoubtedly demonstrated by their
multiwavelength spectrum. On the other hand, the acceleration of nuclei is an elusive phenomenon,
which, from the observational point of view, so far is supported mainly by the impact of efficient
ion acceleration on the thermohydrodynamics of the shockwaves, as inferred from
X-ray observations, and
from the spectral shape of \g-ray emission \cite[e.g.][]{helder2012}. 

The isotopic abundances measured for CRs indicate that $\sim 20\%$ of the material is
synthesized by Wolf-Rayet (WR) stars 100~kyr before the acceleration
\cite{binns2007}. WR
stars represent an evolutionary stage of massive OB stars, and the latter cluster in space and
time within the so-called OB associations. Additionally, $\sim 80\%$ of SNRs
originate from the
gravitational collapse of a massive-star core, therefore often inside an
OB association. All these considerations have given credence to the
hypothesis that at least
part of the CRs are accelerated by the repeated action of shockwaves from
massive stellar winds
and SNRs inside massive stellar clusters.

The propagation of high-energy CRs in the interstellar space is often described in terms of
diffusion by magnetohydrodynamic turbulence with the possible inclusion of convection and
reacceleration \cite{strong2007}. While propagating, CRs loose their energy through many different
interaction processes. Electromagnetic radiation arising from such interactions is the only probe we
have of CR properties in the Galaxy beyond direct measurements performed in the
solar system.
Interstellar \g-ray emission is produced by CR interactions with the
interstellar medium, via nucleon-nucleon inelastic
collisions and electron Bremsstrahlung, and with the low-energy interstellar
radiation fields via
inverse-Compton (IC) scattering by electrons.

The intermediate steps, i.e. the escape of CRs from their sources and the early propagation in the
surrounding medium, have escaped observations so far. Massive stars generate intense radiation,
powerful winds and explosions which alter their neighborhood and power fast
shockwaves and
strong supersonic turbulence. If part of the CRs are accelerated in regions of massive-star
formation, this turbulent medium influences their evolution, e.g. through
confinement and reacceleration phenomena counter-balanced by increased radiative
losses.

The Cygnus~X region is home to numerous massive stellar clusters \cite{leduigou2002}
embedded in a complex of giant molecular clouds. Most of the molecular gas along the line of sight
shows interactions with the stellar clusters, notably with Cyg~OB2 at a
distance of $\sim
1.4$~kpc \cite{negueruela2008}. At a comparable distance we find the SNR \g~Cygni,
e.g. \cite{ladouceur2008}, which appears as a potential CR accelerator. In addition, Milagro
detected very high-energy \g-ray emission from the complex \cite{abdo2007}, pointing to the presence
of enhanced CR densities.

In this paper we summarize the results of our analysis of two years of high-energy \g-ray
observations of the Cygnus complex by the \F{} Large Area Telescope (LAT) \cite{atwood2009}.
The results, discussed in detail in \cite{LATcocoon2011,LATCygISM2012}, reveal an unexpected
scenario, where the low-density cavities blown by the massive stellar clusters
form a cocoon for freshly-accelerated particles, whereas the CR population
averaged over the whole
complex is similar to that found in other more quiet segments of the Local
Arm.

\section{Data and Analysis}

We analyzed the region at Galactic longitudes $72^\circ
\leq l \leq 88^\circ$ and latitudes $-15^\circ b +15^\circ$. The multiwavelength
data and analysis method are detailed in \cite{LATCygISM2012}: in
this section we summarize the main features. 

We have used \g-ray data from the first two years of observations by the \F~LAT between 100~MeV and
100~GeV, applying tight rejection criteria to reduce the backgrounds due to charged CRs and the
Earth's atmospheric emission. Below 1 GeV we accepted only \g-rays that
converted into pairs in the
thin section of the LAT tracker, to reduce the angular resolution degradation due to multiple
Coulomb scattering.

The model described hereafter was fitted to the LAT data by a binned maximum likelihood procedure
on a $0.125^\circ \times 0.125^\circ$ angular grid and independently for 10 energy bins spanning
the whole energy range considered, using post-launch LAT instrument response functions.

Since the densities of the bulk of Galactic CRs are expected to vary only on scales larger than
those of the gas clouds, we modeled interstellar \g-ray emission from their interactions as a
linear combination of components traced by atomic
hydrogen ($\hi$) and CO (as a surrogate of molecular hydrogen, $\hd$) split along the line the sight
to separate the Cygnus complex from the more distant
Perseus and outer spiral arms, plus visual extinction residuals to take into
account the
neutral interstellar gas missed by these two tracers. We added an IC model calculated using the
GALPROP code \cite{strong2007} and an isotropic component to account for the extragalactic diffuse
\g-ray background and the residual background due to charged CR interactions in the LAT.

Below a few GeV the \g-ray emission from the region is dominated by the three
bright pulsars
J$2021+3651$, J$2021+4026$ and J$2032+4127$. To be more sensitive to
interstellar
structures, at energies $<10$~GeV their emission was dimmed by excluding from the analysis photons
detected in their periodic pulses in a circular region around each pulsar
with a radius
comparable to the LAT point-spread function (PSF) 95\%
containment as a function
of energy.

Other sources from the 1FGL Catalog \cite{abdo20101FGL} either identified with, or associated with
high-confidence to sources known from other wavelengths were included as well in the analysis
model.  We modeled \g-ray extended emission associated to the Cygnus Loop SNR as described in
\cite{katagiri2011}.

We detected significant \g-ray emission associated with the radio shell of the
\g~Cygni SNR and the very high-energy \g-ray source VER~J2019+407 located on its rim. They were
modeled using geometrical templates, namely a $0.5^\circ$ disk for the radio shell of the SNR
\cite{green2004} and a Gaussian for VER~J2019+407 \cite{weinstein2009}.

\begin{figure*}[!htb]
\begin{center}
 \includegraphics[width=1.\textwidth]{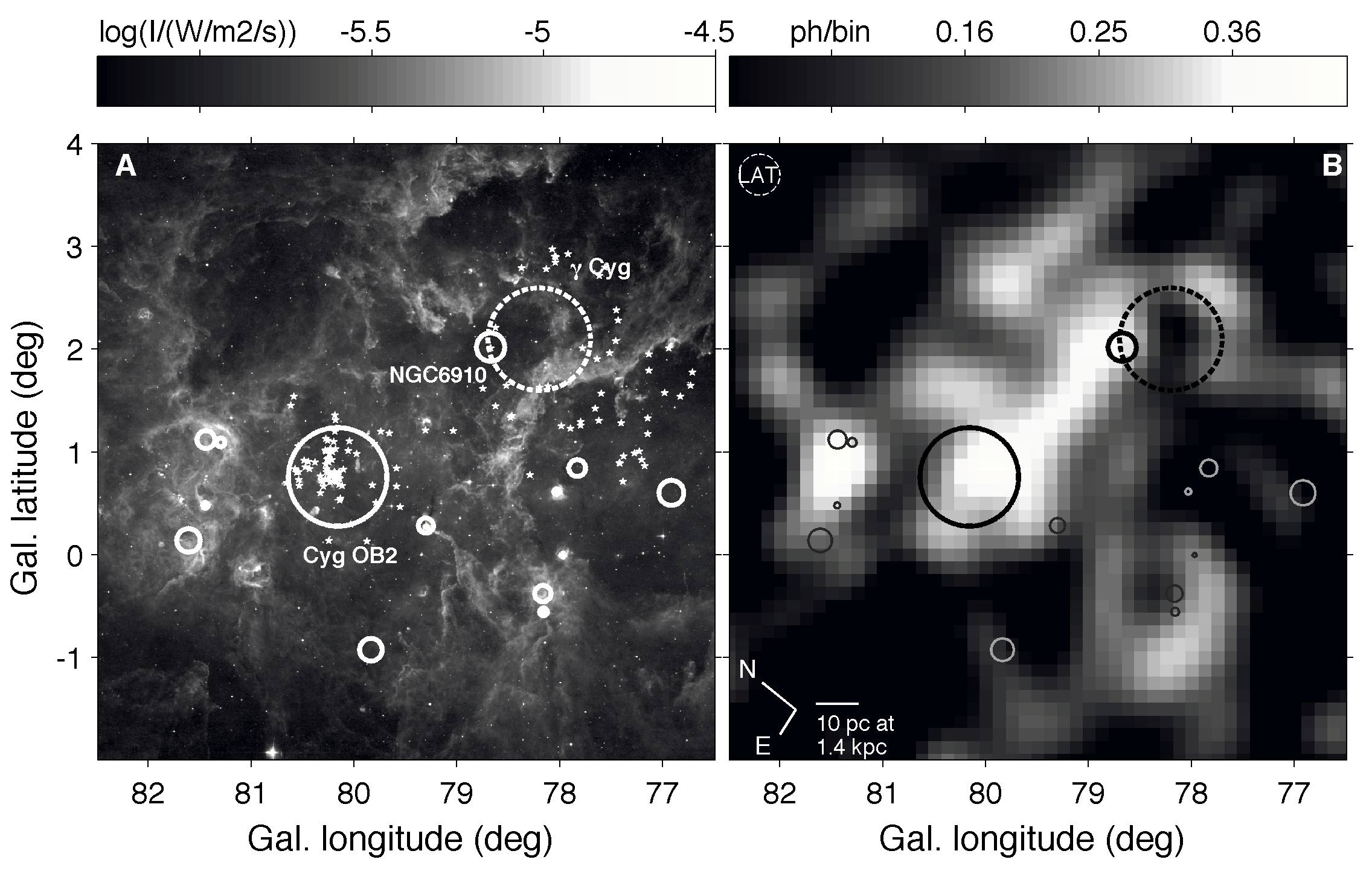}
\caption{A: 8~$\upmu$m map of the Cygnus X region (MSX) showing emission from heated dust
grains; massive OB stars (stars) and associations (circles) are
outlined; the dashed circle shows the rim of the $\gamma$~Cygni SNR; the brightest
features in the map correspond to the photon-dominated regions delimiting the walls of the
ionized cavities carved in the surrounding dense clouds by the intense UV light.
B: \g-ray excess map at energies
$>10$~GeV (LAT, see text for details about the analysis). The LAT
68\%
point-spread function (PSF) is shown by the dashed circle in the upper corner.
Edited from \cite{LATcocoon2011}.}\label{morph}
\end{center}
\end{figure*}

Residuals remain unaccounted for after fitting this model, hereafter the background model,
to the LAT data (Fig.~\ref{morph}B).
The residuals peak toward the central star-forming region of Cygnus X. We
modeled the excess as a Gaussian, obtaining as best-fit parameters
$(l,b)=(79.6^\circ\pm0.3^\circ,1.4^\circ\pm0.4^\circ)$ for the centroid and a radius of
$2.0^\circ\pm0.2^\circ$, with an improvement in the fit to the data above 1 GeV at the $\sim
10\sigma$ level. The hypothesis of extended emission
provides a better fit to the data than an ensemble of discrete point
sources. No spectral variations across the emission region could
be detected \cite{LATcocoon2011}. 

The residual emission (Fig.~\ref{morph}) extends from the Cyg~OB2 stellar
cluster to the \g~Cygni
SNR, peaking toward the most conspicuous clusters like Cyg~OB2 itself and
NGC~6910. However, it is more broadly distributed than any of these
objects. It is unlikely
that the broad \g-ray excess is produced by a wind nebula powered by
PSR J2021+4026 (likely to be associated with the 7-kyr old \g~Cygni SNR) or by PSR J2032+4127 (which
is probably located behind Cygnus in the Perseus arm, \cite{abdo2010psrcat}). We
cannot rule out the presence of a wind nebula powered by a yet undiscovered pulsar, but the
morphology of the \g-ray excess strongly suggests an interstellar origin. Indeed, the \g-ray excess
is  bounded by the photon-dominated regions, visible through
dust emission at 8~$\upmu$m. The excess closely follows the morphology of the
ionized cavities formed by the boisterous winds of the numerous
massive star clusters in Cygnus X, as in a cocoon.

\section{A Cocoon of Freshly Accelerated Cosmic Rays in Cygnus X} 

The emission from the cocoon is hard and extends to 100~GeV. The cocoon
overlaps the TeV source
MGRO~J2031+41 detected by Milagro. Those overlap as well the arcmin source TeV
J2032+4130 detected by HEGRA and MAGIC. However, the flux measured by Milagro
above 10 TeV exceeds the extrapolation of that measured by HEGRA at lower
energies \cite{abdo2007}.

Ionized gas was not part of the background model. It can be traced by
free-free emission in the microwave band. Adding an ionized gas template instead
of the $2^\circ$
Gaussian provides an equally good fit, but only at the expense of a hard spectrum, much harder
than for the other gas phases (\S~\ref{CRcontent}). It is consistent with that
obtained with the Gaussian template \cite[][Fig.~S7]{LATcocoon2011}.

\begin{figure}[!htb]
 \begin{center}
  \includegraphics[width=0.5\textwidth]{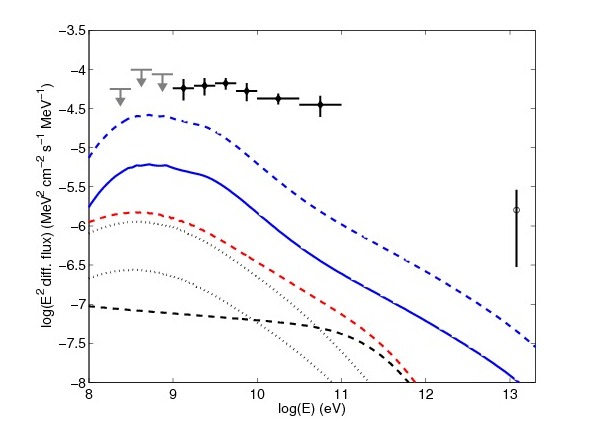}
\caption{Spectral energy distribution of the cocoon (filled circles). Below 1 GeV 95\%
confidence-level upper limits. The empty circle gives the flux of MGRO~J2031+41
subtracted by the flux of the compact source TeV J2032+4130 extrapolated at
energies $>10$~TeV
\cite{abdo2007}. The continuous (blue) line and upper dashed (blue) line show the
spectrum expected from ionized gas illuminated by CRs with the local interstellar spectrum,
assuming an effective electron density of $n_\mathrm{eff}=10$~cm$^{-3}$ and
2~cm$^{-3}$, respectively, to calculate the ionized gas densities from free-free
emission intensities. The intermediate (red) dashed line shows the total IC expected if
CRs had the local interstellar spectrum across the Cygnus X region, including emission
from the tow major stellar clusters Cyg~OB2 (upper --black-- dotted line) and
NGC~6910 (lower --black--
dotted line), as well as from the enhanced optical and infrared radiation fields
pervading the Cygnus~X cavities (lower dashed line). From \cite{LATcocoon2011}.}\label{spec}
 \end{center}
\end{figure}

In Fig.~\ref{spec} we compare
the \g-ray spectrum of the cocoon with predictions based on the \emph{local} CR spectrum, i.e. the
CR spectrum in the interstellar space near the Earth inferred from direct CR
measurements,
\g-ray observations of nearby
clouds and radio observations of synchrotron emission produced by CR electrons
and positrons.
Such predictions fail to reproduce the observed \g-ray spectrum. If we assume that the emission is
purely hadronic we need to amplify the \emph{local} proton and helium spectra by
$(1.6-1.8)\times(E/10\,\mathrm{GeV})^{0.3}$. If we assume that it is purely
leptonic we need an
amplification factor of $\sim 60 \times (E/10\,\mathrm{GeV})^{0.5}$ for the
\emph{local} CR electron and positron
spectrum. The hardness of the \g-ray spectrum points to freshly-accelerated
particles since the
lifetime of TeV electrons in the enhanced radiation environment of the cocoon is $<20$~kyr and the
escape time for $> 0.1$~TeV nuclei is $< 50$~kyr for the average Galactic diffusion coefficient.

The \g~Cygni SNR is one of the potential sites where the young CRs might have
been accelerated. The
shockwave still shelters energetic particles shining in \g-rays. CR acceleration
models (taking into
account CR pressure and magnetic feedback) show that, based on the shockwave properties
inferred from optical and X-ray observations, \g~Cygni may have released protons up to 80--300~TeV
and electrons up to 6--30~TeV at the end of the the free-expansion phase approximately 5 kyrs ago.
If the dominant particle transport mechanism were diffusion, and the diffusion coefficient were
similar to that in the Galaxy at large, particles released by \g~Cygni at that
time could fill
the whole cocoon. Yet, there is no proof of a physical link between the SNR and the Cygnus~X
cavities. Moreover, the anisotropy of the particle release
with respect to the remnant (Fig.~\ref{morph}), as well as the
comparable extension of the \g-ray emission at different energies, would point
to an advection-dominated scenario instead of diffusive transport, but there is
no proof either that the shockwave breaks free on its western rim
adjacent to the cocoon. In the absence of advection the short diffusion lengths
expected from the
high magnetic turbulence level in the Cygnus~X region could rule out the young \g~Cygni
remnant as the only source of energetic particles in the cocoon.

Superbubbles created by the collective action of SNR and stellar wind shockwaves are considered as
potential CR accelerators \cite[e.g.][]{bykov2001,ferrand2010}. The few Myr old clusters in
Cygnus~X have had time to produce very few SNRs, if any. The total \g-ray
luminosity of the
cocoon above 1~GeV, $(9\pm2)\times 10^{27}W$ at 1.4~kpc, however, represents only $\sim 0.03\%$ and
$\sim 7\%$ of the mechanical power injected by stellar winds in Cygnus~OB2 and NGC~6910,
respectively. Wind-powered turbulence in the high-pressure gas within the cocoon, with a typical
energy-containing scale of 10~pc, can efficiently confine and accelerate particles. With
near-saturation turbulence and a regular magnetic field of 2~nT (in pressure balance with the
gas) the particle energy distribution would peak at 10--100~GeV and potentially extend up to
150~TeV. For a typical dimension of $\sim 50$~pc the turbulence would scatter
TeV particles for more than 100~kyr, in agreement with the timescale
inferred from the lack of $^{59}$Ni revealed by direct CR
measurements \cite{binns2007}.

\section{The Cosmic Ray Content of the Cygnus Complex}\label{CRcontent}

A \g-ray excess revealed the presence of a
cocoon of freshly-accelerated CRs over a scale of
$\sim 50$~pc in the heart of Cygnus~X. The average CR
population of the whole Cygnus interstellar complex, over a
scale of $\sim 400$~pc, was probed by the $\hi$ emissivity, i.e. the \g-ray
emission rate per
hydrogen atom, which is derived in the fit of the background model to the LAT
data.

\begin{figure}[!htb]
 \begin{center}
  \includegraphics[width=0.5\textwidth]{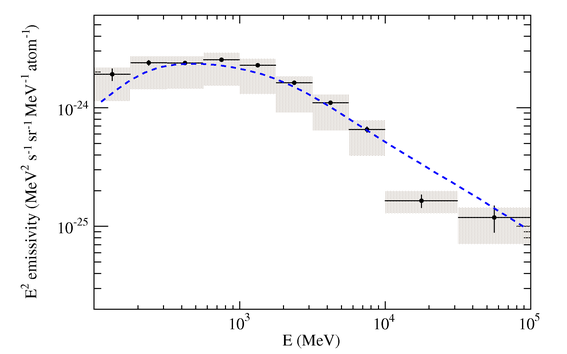}
\caption{\g-ray emissivity per hydrogen atom averaged over the whole
interstellar cloud complex of Cygnus.
Points include statistical uncertainties only, whereas the dashed bands show the
systematic uncertainties related to the $\hi$ column densities and to the LAT effective
area. The dashed (blue) curve shows the model for the \emph{local} interstellar
spectrum
consistent with LAT observations of nearby interstellar clouds
\cite{abdo2009lochiemiss}. From \cite{LATCygISM2012}.}\label{hispec}
 \end{center}
\end{figure}

In Fig.~\ref{hispec} we compare the $\hi$ emissivity spectrum of the Cygnus
complex with the model
for the \emph{local} interstellar spectrum that was
shown to be in
good agreement with LAT measurements of diffuse \g-ray emission from gas within
1~kpc from the Sun \cite{abdo2009lochiemiss}.
LAT data indicate that the CR population averaged over the whole Cygnus complex is similar within
$\sim 20\%$ to that in the much more quiet solar neighborhood (the old
Galactic CR
population). No counterpart to the broad
excess of \g-ray emission seen at
energies $>10$~TeV at $65^\circ \leq l \leq
85^\circ$ \citep{abdo2007,abdo2008milagrodiff} was detected by the LAT so far.
Only the molecular clouds closer to the central star-forming region show
a hint of a younger
CR population from a slight spectral hardening at energies $>10$~GeV
\cite[][Fig.~12]{LATCygISM2012}. This points again to an
efficient confinement of the particles inside the cocoon.

\begin{figure}[!htb]
 \begin{center}
  \includegraphics[width=0.5\textwidth]{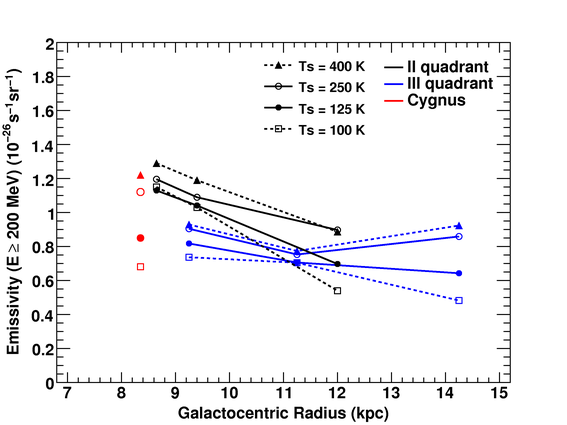}
\caption{\g-ray emissivity of atomic hydrogen at energies $>200$~MeV measured by the LAT in the
Cygnus region (red or light gray), the second Galactic quadrant
\cite{abdo2010cascep} (black) and the third Galactic
quadrant \cite{ackermann20113quad} (blue or gray). Results are shown for
different values of
the $\hi$ spin temperature $T_S$, i.e. the parameter used to determine from radio observations the
column densities of atomic hydrogen, which are currently the main source of uncertainty in this
kind of measurement.}\label{largescalefig}
 \end{center}
\end{figure}

In Fig.~\ref{largescalefig} we broaden the view on the Galaxy at large by
comparing the emissivity
measurement in Cygnus with those in other regions of the local and outer Galaxy. 
In spite of the CR acceleration ongoing in Cygnus and of the steep decline in
number of putative CR sources
toward the outer Galaxy, there are no sizable variations of the CR densities
traced by the
interstellar \g-ray emission \cite[see also][]{LATdiffpapII}.

Unless we invoke the presence of large amounts of missing gas in the outer Galaxy, models assuming
a uniform CR diffusion throughout the Galaxy would require a very large propagation halo or the
presence of unknown CR sources in the outer Galaxy to reproduce the LAT data
\cite{ackermann20113quad}. On the other hand, the strong
turbulence associated with the stellar clusters and CR acceleration
can increase the
diffusion coefficient perpendicular to the Galactic disk, hence make the escape
of CRs faster and wash out the CR density around their
sources \cite{evoli2012}.

\section{Summary}

The \F~LAT observations of Cygnus have added a few more pieces to the
puzzle of Galactic CRs. The discovery of a cocoon of freshly-accelerated CRs in
Cygnus~X confirms the long-standing hypothesis that massive-star forming
regions house particle accelerators. It provides an observational test case
to study the escape of CRs from their sources and the impact of wind-powered
turbulence on their early evolution. It also sheds a new light on the
observation of TeV \g-ray emission from regions harboring massive stellar
clusters
\cite[e.g.][]{aharonian2006diff,aharonian2007westerlund2,
abramowski2012westerlund1}.

The similarity of the CR population averaged over the whole Cygnus complex to
that in the rest of the Local Arm and in the outer Galaxy strengthens the
discrepancy found between LAT observations and propagation models
that predict a strong coupling over a scale of $\sim 1-2$~kpc between CR
source densities and CR
fluxes. Studies of \F~\g-ray observations of
massive star-forming complexes in the inner Galaxy are required to further
investigate this issue.

Some of the usual assumptions concerning the injection and transport of CRs in
the interstellar space are challenged by the current observations.
Better understanding the propagation and interactions of CRs
has important implications for the physics of the interstellar medium, hence for
star formation, in the
Milky Way, as well as in other galaxies including starbusts. It is also a
crucial
prerequisite to look for any tracks of still unknown physical processes hidden
in the sea of energetic particles filling the Universe, such as the products of
decay or annihilation of dark-matter particles. All in all, the next
century of CR Physics looks as exciting as the past one.

\vspace{24pt}
The \textit{Fermi} LAT Collaboration acknowledges support from a number of agencies and institutes
for both development and the operation of the LAT as well as scientific data analysis. These include
NASA and DOE in the United States, CEA/Irfu and IN2P3/CNRS in France, ASI and INFN in Italy, MEXT,
KEK, and JAXA in Japan, and the K.~A.~Wallenberg Foundation, the Swedish Research Council and the
National Space Board in Sweden. Additional support from INAF in Italy and CNES in France for science
analysis during the operations phase is also gratefully acknowledged.

\end{document}